\documentclass[a4paper,USenglish,cleveref, autoref]{lipics-v2021}
\hideLIPIcs  

\nolinenumbers

\usepackage[dvipsnames]{xcolor}
\usepackage{graphicx}
\usepackage{cite}
\usepackage{wrapfig}

\title{Survey of Persistent Memory Correctness Conditions}

\author{Naama Ben-David}{VMware Research, USA}{bendavidn@vmware.com}{}{}
 
\author{Michal Friedman}{ETH, Switzerland}{michal.friedman@inf.ethz.ch}{}{}

\author{Yuanhao Wei}{Carnegie Mellon University, USA}{yuanhao1@cs.cmu.ed}{}{}

\authorrunning{N. Ben-David, M. Friedman and Y. Wei} 

\ccsdesc[500]{Theory of computation~Concurrency}
\ccsdesc[500]{Hardware~Robustness}

\keywords{Persistence, NVRAM, Correctness, Concurrency} 

\newcommand{\Naama}[1]{}
\newcommand{\Michal}[1]{}
\newcommand{\Hao}[1]{}
\newcommand{\maybeDel}[1]{}

\begin{document}

\maketitle

\begin{abstract}
The study of concurrent persistent programs has seen a surge of activity in recent years due to the introduction of non-volatile random access memories (NVRAM), yielding many models and correctness notions that are difficult to compare. 
In this paper, we survey existing correctness properties for this setting, placing them into the same context and comparing them. We present a hierarchy of these persistence properties based on the generality of the histories they deem correct, and show how this hierarchy shifts based on different model assumptions. 
\end{abstract}

\section{Introduction}

Non-Volatile Random Access Memory (NVRAM) is a new type of memory technology that has recently hit the market. Its key feature is that it is \emph{persistent}, like SSDs, but is fast and byte-addressable, much like DRAM. This presents a huge paradigm shift from the way persistence could be achieved in the past; techniques that worked well for sequential block-granularity storage cannot be efficiently used with NVRAMs. Achieving persistence with NVRAM has the potential to speed up applications by orders of magnitude.

However, before designing persistent algorithms for NVRAM, we must first answer a more basic question:
    What does it mean for an algorithm to be persistent?

Despite algorithms relying on external storage for persistence for decades, the answer to the above question is not clear in the context of faster, byte addressible NVRAM.
Due to the slow nature of external storage, persistence was achieved via periodic checkpointing that allowed considerable progress to be lost upon a system crash. However, with the introduction of NVRAM, we now have new possibilities; its speed and byte addressability allow us to demand stronger persistence guarantees. 
In particular, it is now realistic to require that virtually \emph{no progress be lost} upon a crash, and that a program be able to continue where it left off upon recovery.

The above requirement, while appealing, is in fact not very precise. Due to registers and caches remaining volatile, individual instructions and memory accesses are applied to volatile memory first, and are then persisted separately. If a system crash occurs between when an instruction is executed and when its effect is applied to NVRAM, progress will inevitably be lost. However, it is possible to define how much progress it is okay to lose, and at what point in the execution we expect each instruction's effect to be persisted. For example, we can ensure no completed operation will be lost upon a crash.

Furthermore, aside from reliably persisting the effect of operations, we must also consider what it means to recover and continue an execution after a system crash. For example, do we execute a synchronous recovery protocol, or let each process recover individually? Do we expect executing processes to pick up where they left off in their program? How can we ensure that this can be done? These questions and others have been the focus of several works in recent years, aiming to characterize the desirable and achievable persistence properties of NVRAM programs.

In this survey, we discuss definitions of persistence that exist in the literature. As this is an actively and quickly developing field of study, there are many different notations, terminologies, and definitions that often refer to similar notions. We put these definitions into the same terminology, and compare them to each other. 
Using this point of view, we arrange the definitions into a hierarchy, based on the set of execution histories that satisfies every definition. Interestingly, this hierarchy changes depending on specific model assumptions made. We outline common model assumptions and illustrate their effect on these definitions. 
Finally, we also introduce new definitions that are intuitively `missing' from the literature so far, and explain how they fit into the existing persistence definition hierarchy.

We note that this survey is meant to make sense of the various persistence definitions and to guide researchers and algorithm designers when choosing which model and definition to adopt. However, this survey does not cover the many different algorithms, techniques, and applications that have been developed for NVRAM programming in recent years.
\section{Model Assumptions}
We begin by describing the model assumptions that are common to all persistence notions we will discuss, along with common terminology we will be using to discuss the persistence definitions that appear in the literature. In Section~\ref{sec:modelflex}, we outline common differences among existing persistence models that have a significant affect on the meaning of these definitions.

\subsection{Common Model}\label{sec:commonmodel}

We consider a system of $n$ asynchronous processes $p_1 \ldots p_n$. Processes may access \emph{shared base objects} with atomic \emph{read, write, and read-modify-write} primitives. Each process also has access to \emph{local variables} that are not shared with any other process. Objects (both local and shared) may be \emph{volatile} or \emph{non-volatile}, which affects their behavior upon a crash. 
This behavior is specified differently in different models in the literature (as described in Section~\ref{sec:modelflex}). 
More complicated objects may be implemented from base objects.

\textbf{Histories and events.}
A \maybeDel{ \emph{history} is a sequence of accesses to base objects. An \emph{interpreted} or \emph{high-level}} history is defined in terms of \emph{events} as follows.
There are three types of events: an \emph{invocation} event $obj.inv_i(op, v)$ which invokes operation $op$ on object $obj$ by process $p_i$ with argument $v$, a \emph{response} event $obj.res_i(op, v)$ in which $obj$ responds to $p_i$'s invocation on $op$ with return value $v$, and \emph{crash} events. Depending on the specific NVRAM model, a crash may be associated with an individual process, or may be system-wide, meaning that it is associated with all processes. A crash resets all the volatile local. 

\textbf{Properties of Histories.}
A response $res$ is said to \emph{match} an invocation $inv$ in $H$ if $obj$, $op$, and $i$ are the same for both, and $res$ is the next event in $H\vert p$ after $inv$. An operation is said to be \emph{complete} in $H$ if both its invocation and a matching response appear in $H$. Otherwise, if an operation was invoked but was not completed, the operation is said to be \emph{pending}.
Given a history $H$ and two operations $op_1$ and $op_2$, $op_1$ \emph{happens before} $op_2$ if $res$ of $op_1$ precedes $inv$ of $op_2$ in $H$, and is denoted by $op_1<_Hop_2$. Operations that do not preserve this order are said to be \emph{concurrent}.
A history $H$ \emph{projected on process $p$}, denoted $H\vert p$, is the sub-history of $H$ composed of exactly all of the events in $H$ associated with $p$, as well as any system-wide crash events.
Similarly, a history $H$ \emph{projected on object $obj$}, denoted $H\vert obj$, is the sub-history of $H$ composed of exactly all of the events in $H$ associated with $obj$, as well as any crash events.
A history is said to be \emph{sequential} if there is an immediate matching response after every invocation, except possibly that last. Two histories $H$ and $H'$ are said to be \emph{equivalent} if for every process $p$, $H\vert p=H'\vert p$.
Given the happens before relation, a \emph{$<_H$-consistent cut} of $H$ is a sub-history $H_s$, such that if event $e_2\in H_s$ and $e_1<_He_2$ then $e_1\in H_s$.
A set of histories $S$ is \emph{prefix-closed} if for every sequential history $H$, its prefix also belongs to $S$. A \emph{sequential specification} for an object $obj$ is a prefix-closed set of all allowed sequential histories that contains only operations over $obj$. 
A sequential history $H$ is \emph{legal} if each history projected on object $obj$, $H\vert obj$, belongs to the sequential specification of $obj$.
A history $H$ is \emph{well-formed} if for every process $p$, every invocation in $H\vert p$ is immediately followed by either a matching response or a crash, and every response in $H\vert p$ is immediately preceded by a matching invocation.
Another useful and important property is called \emph{locality}. Given a property $P$ and a history $H$, $P$ is said to be \emph{local} if given a history $H$ in which, for every object $O$, $H \vert O$ satisfies $P$, $H$ also satisfies $P$.

\subsection{Model's Flexibility}\label{sec:modelflex}

Models in which crashes are taken into account may have different assumptions regarding if and when crashed processes must recover. The varied assumptions are described below.

    \textbf{System-wide vs. individual crashes.}
    Depending on the model, a crash event can be associated with all processes or with just one. Both versions have been considered in the literature. The model of individual process crashes~\cite{ben2019delay,aguilera2003strict,attiya2018nesting,blelloch2018parallel} is more general than those that consider system-wide crashes~\cite{izraelevitz2016linearizability,berryhill2016robust, friedman2018persistent}. 
    However, system-wide crashes better represent power outages, which are the primary faults that may be experienced on real machines.
    
    \textbf{Same processes vs. new processes after a crash.}
    In some models~\cite{izraelevitz2016linearizability}, processes do not recover after a crash event, and new processes are spawned to continue the execution. This is a simplifying assumption over the model considered in other works~\cite{attiya2018nesting,golab2019recoverable,guerraoui2004robust,berryhill2016robust,ben2019delay,blelloch2018parallel}, in which a process may recover and continue its execution after a crash. 
    The assumption that processes do not recover is more in line with the classic shared memory model. However, depending on system requirements, it might not suffice; upon a system crash, we might want to resume executing the same processes from the moment they crashed and provide the ability to act as if a crash has never happened.


    \textbf{Volatile vs. non-volatile shared memory.}
    Some works~\cite{ben2019delay,attiya2018nesting,berryhill2016robust} assume that there are no volatile shared objects. That is, all communication between processes happens on persistent memory, and is unaffected by crash events. In these models, the challenge is how to handle the loss of local variables.
    Other works~\cite{izraelevitz2016linearizability,friedman2018persistent,correia2018romulus,friedman2020nvtraverse,friedman2021mirror,wen2020montage} assume that all shared objects are volatile, but have a non-volatile copy that may occasionally be updated.
    This assumption is usually accompanied by the assumption that crash events are system-wide, and the state of volatile shared memory is lost upon a crash. 
    In these models, it is assumed that there is a mechanism to persist values in volatile memory (shared or local) explicitly through low-level instructions called \emph{flush} and \emph{fence}.
    Furthermore, motivated by automatic cache evictions, it is often assumed that some parts of the shared memory may be \emph{implicitly} persisted at unknown times. 
    This further complicates the task of maintaining a consistent state in non-volatile memory to use upon recovery. This assumption stems from the way that modern machines work, where variables are shared on a volatile cache that may automatically evict some cache lines to non-volatile memory.

\subsection{A Running Example}

\begin{figure}[htp]
    \centering
    \includegraphics[width=10cm]{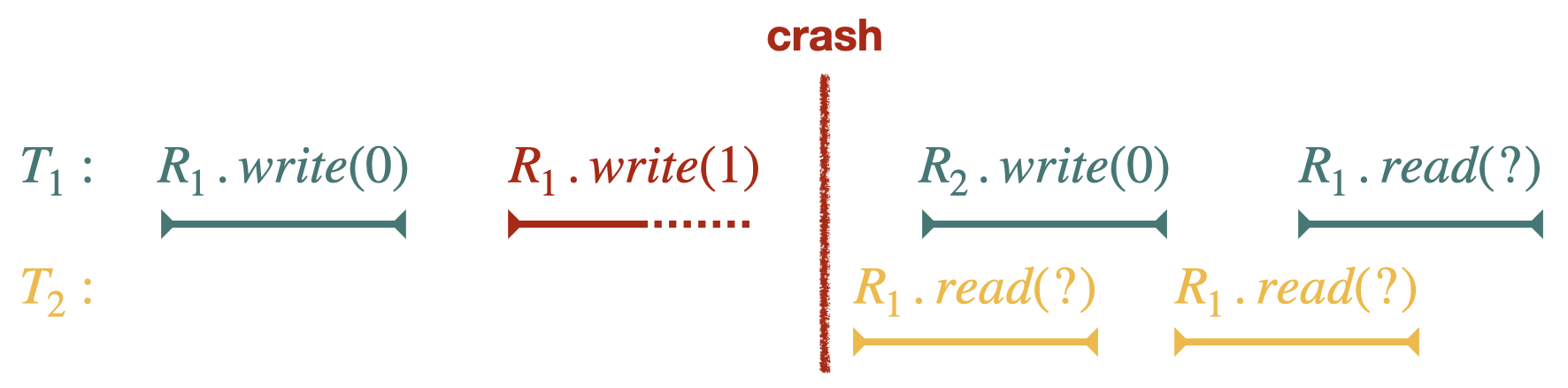}
    \caption{An example: Two threads reading and writing to two registers.}
    \label{fig:runningexample}
\end{figure}

Throughout the entire survey, we use variations of a simple example showing two threads, $T_1$ and $T_2$, executing their instructions on two multi-reader multi-writer registers, $R_1$ and $R_2$. Depending on the explained definition, we show when an interrupted invocation of a write operation by a crash can complete and take effect. In addition, we reason about the returned values of the threads that read the register whose the write to was interrupted by the crash. The example is shown in Figure~\ref{fig:runningexample}. The thread $T_1$ first writes the value $0$ to register $R_1$. Then, it writes the value $1$ to the same register, which is interrupted by a crash. This write may or may not survive the crash. After the crash, thread $T_1$ writes $0$ to the second register, $R_2$. Finally, it reads the value from register $R_1$. The returned value from the read might be either $0$ or $1$, depending on the definition that the history satisfies. In the meanwhile, thread $T_2$ reads the value from register $R_1$ twice.

\section{Existing Correctness Properties}\label{sec:definitions}
All of the following correctness conditions describe the meaning of being in a consistent state after a crash. 
It is helpful to view these correctness conditions in terms of how they allow operations that remain pending due to crash events to complete.
This approach was initially suggested by Berryhill et al.~\cite{berryhill2016robust}, where the differences among some of the definitions occur in how the completed history is defined. Using this point of view, we can arrange the definitions into a hierarchy, by the set of histories that satisfies every definition. Interestingly, this hierarchy changes depending on the specific model assumptions made.

    \textbf{Linearizability}
    was originally defined by Herlihy and Wing~\cite{herlihy1990linearizability}. Intuitively, it requires that each operation takes effect at a single point in time between its invocation and its response. More formally, to define linearizability, we first define a \emph{completion} of a history.  $H'$ is said to be a completion of a history $H$ if for each pending invocation in $H$, it is either removed from $H'$, or a matching response event is appended to the end of the history in $H'$. 
    A history $H$ is linearizable if there exists a completion $H'$ of $H$ and there exists a legal sequential history $S$ such that:
    \begin{enumerate}
    \item \label{cond:lin1} $H'$ is equivalent the legal sequential history $S$
    \item \label{cond:lin2} $<_{H'} \subseteq <_S$
    \end{enumerate}
    
\begin{figure}[htp]
    \centering
    \includegraphics[width=10cm]{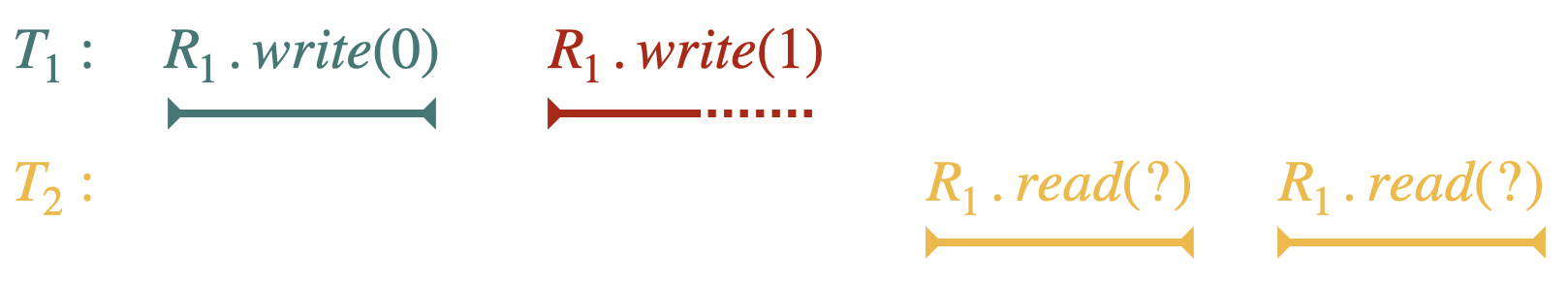}
    \caption{Linearizability example}
    \label{fig:linearizabilityexample}
    \end{figure}

    Linearizability inherently does not consider any crash events, but allows individual processes to stop executing, which may be seen as a single process crash. We adjust Figure~\ref{fig:runningexample} to demonstrate the Linearizability definition by stopping $T_1$ after it invokes the operation that writes $1$ and removing all the subsequent operations. The adjusted example is presented in Figure~\ref{fig:linearizabilityexample}.
    
    
\begin{figure}[htp]
    \centering
    \includegraphics[width=10cm]{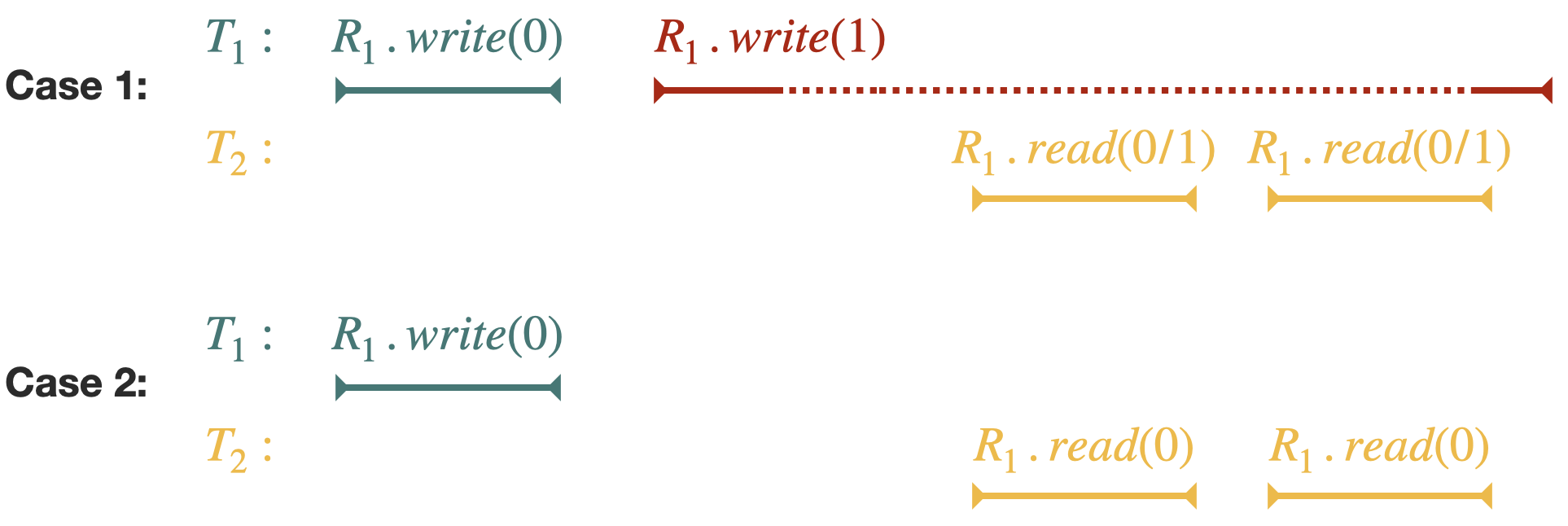}
    \caption{Linearizable history completions}
    \label{fig:linearizabilitycompletions}
\end{figure}

    In this case, we can complete the history by either adding a response to the interrupted invocation at the end of the history or removing it from the history. The cases are demonstrated in Figure~\ref{fig:linearizabilitycompletions}. In the first case, if the operation takes effect before one of the reads of $T_2$ takes effect, then the returned value by that and all the following reads must be $1$. In all the other scenarios, including the second case, where the invocation of the stopped operation is removed from the history, the read that is executed by $T_2$ must return $0$.

    \textbf{Strict Linearizability} was defined by Aguilera and Fr{\o}lund~\cite{aguilera2003strict} in the context of abortable operations. They did not have NVRAM in mind when defining this correctness condition, and considered individual process crashes. We present their definition framed under this model, in a similar fashion as was presented by Berryhill et al.~\cite{berryhill2016robust}. 
    
    The difference between strict linearizability and classic linearizability is in how to handle crashes, 
    which the classic definition does not address. Informally, strict linearizability requires that each operation pending due to a crash either take effect before the crash, or never take effect at all. This requirement is expressed in the definition of a variant of a history completion, as follows. 
    A \emph{strict completion}, $H'$, of a history $H$ is created by, for each pending invocation $inv$ in $H$, either adding a matching response before the first crash event after $inv$ in $H$, or removing $inv$ from the history, and also removing all crash events.
    A history $H$ is then strictly linearizable if there exists a \emph{strict completion} $H'$ that satisfies conditions~\ref{cond:lin1} and~\ref{cond:lin2} from the Linearizability definition.
    Like classic linearizability, strict linearizability is a local property.

    Note that other than the way $H'$ is formed, the definitions of strict linearizability and classic linearizability are the same.
    
\begin{figure}[htp]
    \centering
    \includegraphics[width=10cm]{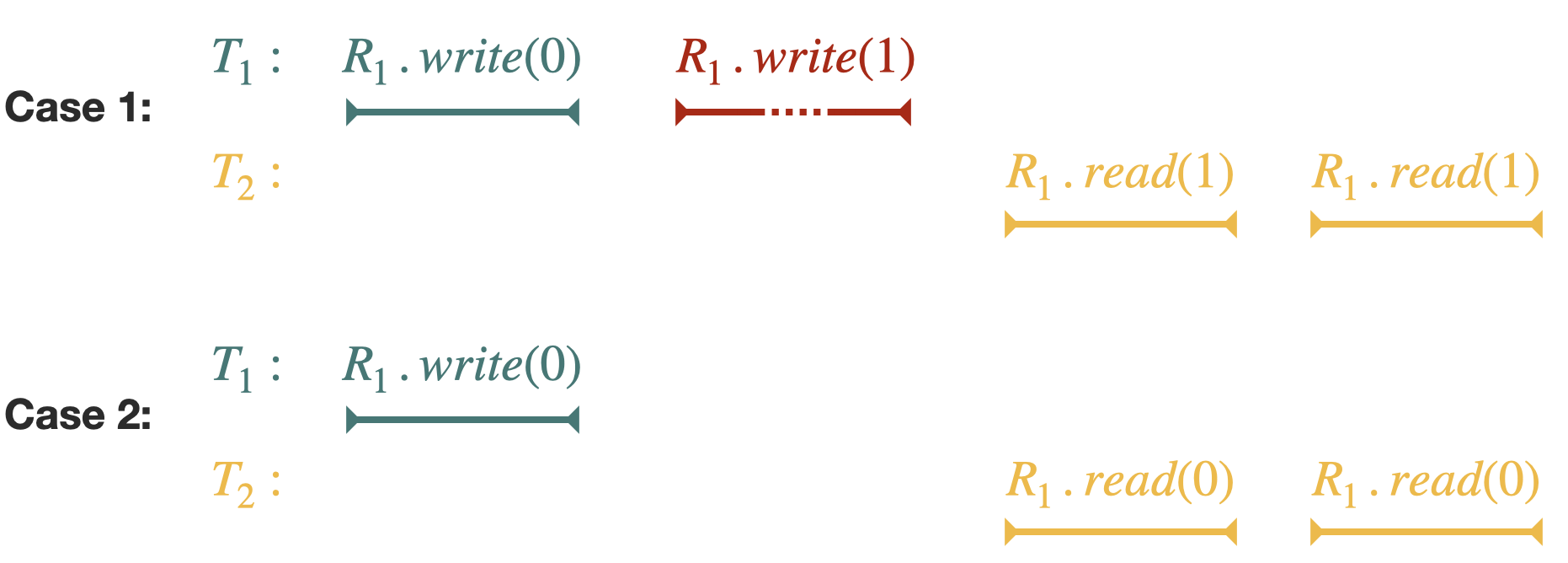}
    \caption{Strict history completions}
    \label{fig:strictcompletions}
\end{figure}

    According to their model, a crashed process cannot continue. Therefore, we adjust Figure~\ref{fig:runningexample} by removing all subsequent operations that follow the crash, and relate the crash only to $T_1$. Figure~\ref{fig:strictcompletions} demonstrates the two possible strict completions. In the first case, the crashed operation is completed before the crash. In this case, $T_2$ must read the value $1$ from register $R_1$. In the second case, the crashed operation is removed from the history, causing $T_2$ to read the value $0$ from $R_1$.
    
    \textbf{Persistent Atomicity} was defined by Guerraoui and Levi~\cite{guerraoui2004robust}. Their model relies on asynchronous message passing, but can easily be converted to a shared memory model. Processes crash individually and must run a recovery procedure when they continue their execution. This definition enables more histories by relaxing the deadline in which the pending operations must complete. Informally, a pending operation that was interrupted by a crash must take effect before the next invocation by the same process on any object.
    Formally, a \emph{persistent completion}, $H'$, of a history $H$ is created by, for each pending invocation $inv$ in $H$, either adding a matching response before the next invocation by the same process in $H$, or removing $inv$ from the history, and also removing all crash events.
    A history $H$ satisfies persistent atomicity if there exists a \emph{persistent completion} $H'$ that satisfies conditions~\ref{cond:lin1} and~\ref{cond:lin2} from the linearizability definition. Despite its generality, it is not local~\cite{berryhill2016robust}.
    
\begin{figure}[htp]
    \centering
    \includegraphics[width=10cm]{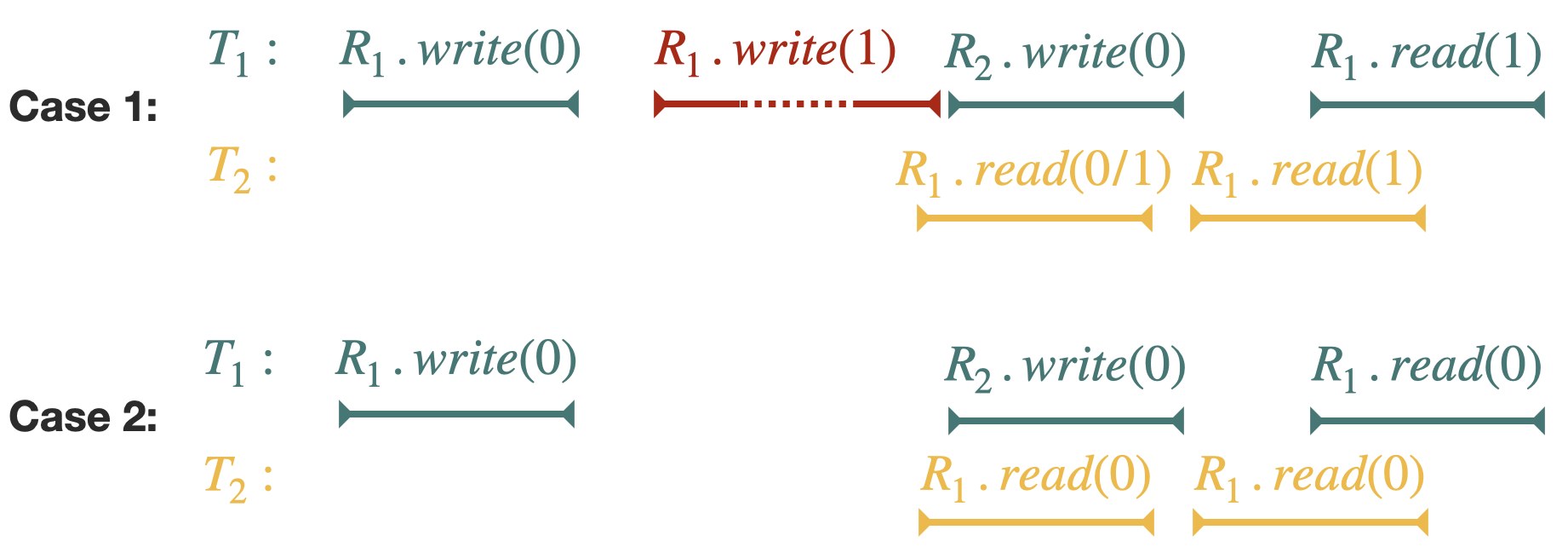}
    \caption{Persistent history completions}
    \label{fig:persistentcompletions}
    \end{figure}

    According to our example, the two possible persistent completions are demonstrated in Figure~\ref{fig:persistentcompletions}. In the first case, the crashed operation is completed before the next invocation executed by $T_1$. In this case, the read executed by $T_1$ from register $R_1$ after the crash must read the value $1$. Same value is read by the second read operation that is executed by $T_2$. Note that the first read of $T_2$ might actually still read the value $0$ if it takes effect before the write of $1$ takes effect.
    In the second case, the crashed operation is removed from the history, causing $T_1$ and $T_2$ to read the value $0$ from $R_1$.
    
    \textbf{Recoverable Linearizability} was defined by Berryhill et al.~\cite{berryhill2016robust}. Its aim is to maintain locality while allowing more flexibility than persistent atomicity and strict linearizability. The model assumes full system crashes, and allows the same processes to continue their execution after a crash without an explicit recovery step. Both volatile and non-volatile regions may exist.
    
    Note that other than the way $H'$ is formed, the definitions of strict linearizability, persistent atomicity, and classic linearizability are the same. Recoverable linearizability deviates from this pattern; it still makes use of history completions, but changes how they are used to determine whether the original history satisfies the definition.
    
    More specifically,
    \emph{recoverable linearizability} uses \emph{strict completions}, defined in the same way as in the strict linearizability definition.

    The difference is reflected in the use of the new following definition, the \emph{invoked before} relation, that was also proposed by Berryhill et al.~\cite{berryhill2016robust}. When operations are applied to the same object by the same process, this definition helps to prevent program order inversion.
     Given a history $H$ and two operations $op_1$ and $op_2$ invoked by the same process $p$ on the same object $O$, $op_1$ is \emph{invoked before} $op_2$ (denoted by $op_1\ll_Hop_2$) if $inv$ of $op_1$ precedes $inv$ of $op_2$ in $H$.

     A history $H$ satisfies recoverable linearizability if there exists a \emph{strict completion} $H'$ and there exists a legal sequential history $S$ such that:
     \begin{enumerate}
         \item \label{cond:rlin1} $H'$ is equivalent to a legal sequential history $S$
         \item \label{cond:rlin2} If $op_1$ happens before $op_2$ in $H$, then $op_1$ happens before $op_2$ in $S$
         \item \label{cond:rlin3} If $op_1$ is invoked before $op_2$ in $H$, then $op_1$ happens before $op_2$ in $S$ \Hao{update proofs}
     \end{enumerate}
     
     By relaxing~\ref{cond:rlin2} and using $H$ instead of $H'$, an operation interrupted by a crash can take effect after the crash.
     Note that this means that the order in which two operations executed by the same process on different objects may differ from the order in which they are invoked.

     Intuitively, the operation that was interrupted by a crash must take effect sometime before the next invocation by the same process on the same object, but not necessarily before the next invocation by the same process on \emph{any} other object. 
     
\begin{figure}[htp]
    \centering
    \includegraphics[width=10cm]{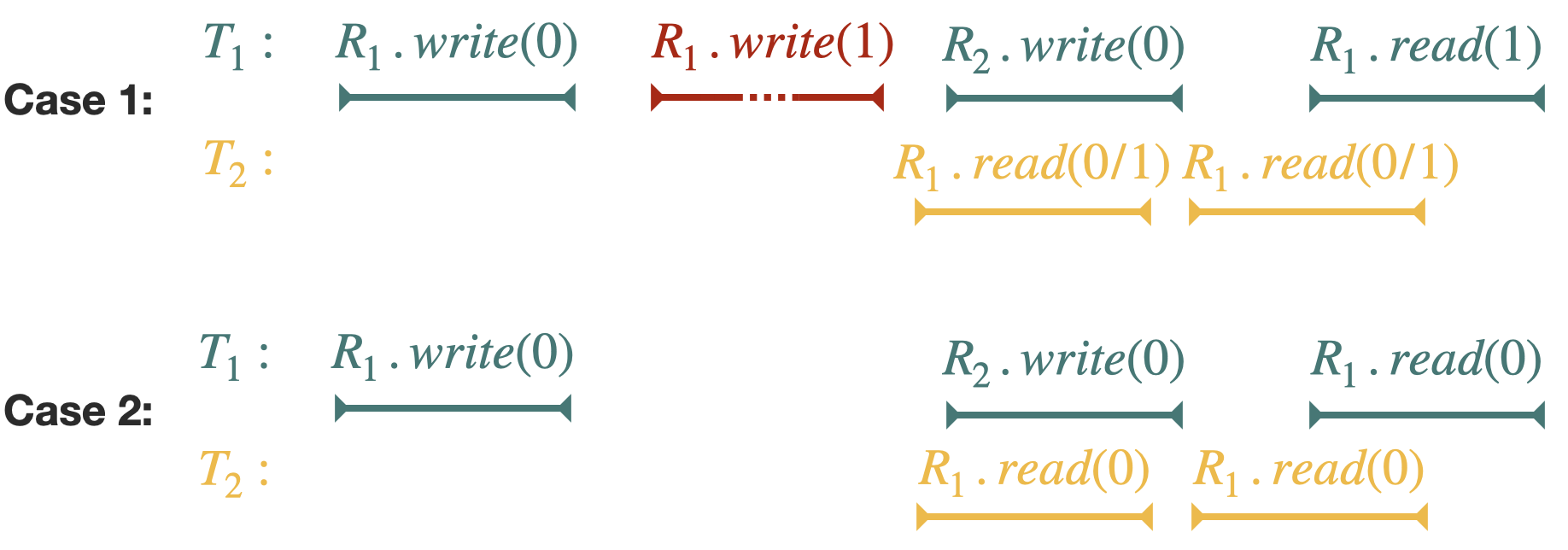}
    \caption{Recoverable history completions}
    \label{fig:recoverablecompletions}
\end{figure}

     For example, consider the running example where the invocation of write(1) that is executed by thread $T_1$ on register $R_1$ and is interrupted by a crash, and another write invocation, write(0), by the same thread on register $R_2$ after the crash. The possible recoverable completions are depicted Figure~\ref{fig:recoverablecompletions}. In the first case, a completion to write(1) was added to the history right before the crash. According to the definition, write(0) may actually take effect before write(1). Therefore, if write(1) takes effect after the reads that are executed by $T_2$, their return value would be $0$. The read that is executed by $T_1$ however, always returns $1$ in this case. In the second case, write(1) is removed from the history and all subsequent reads return $0$.
     
     
    \textbf{Durable Linearizability}, defined by Izraelevitz et al.~\cite{izraelevitz2016linearizability}, simplifies the model by assuming that at a crash, all processes fail, and do not recover~\cite{izraelevitz2016linearizability} -- new processes are spawned instead. Moreover, model assumes full system crash and both volatile and non-volatile regions may exist. Making this change to the model unifies recoverable and persistent linearizability. To see why, recall that recoverable and persistent completions both restrict the placement of matching responses only when the same process invokes a new operation after a crash. If this never happens, then both conditions allow all matching responses to be appended at the end of a history. This unified condition is called \emph{durable linearizability}.
    To put this definition in the same context as others, we define durable linearizability with the following new durable completion definition. A \emph{durable completion}, $H'$, of a history $H$ is created by, for each pending invocation $inv$ in $H$, either adding a matching response at the end of $H$, or removing $inv$ from the history, and also removing all crash events. 
    A history $H$ satisfies durable linearizability if there exists a \emph{durable completion} $H'$ that satisfies conditions~\ref{cond:lin1} and~\ref{cond:lin2} from the linearizability definition.
    
\begin{figure}[htp]
    \centering
    \includegraphics[width=10cm]{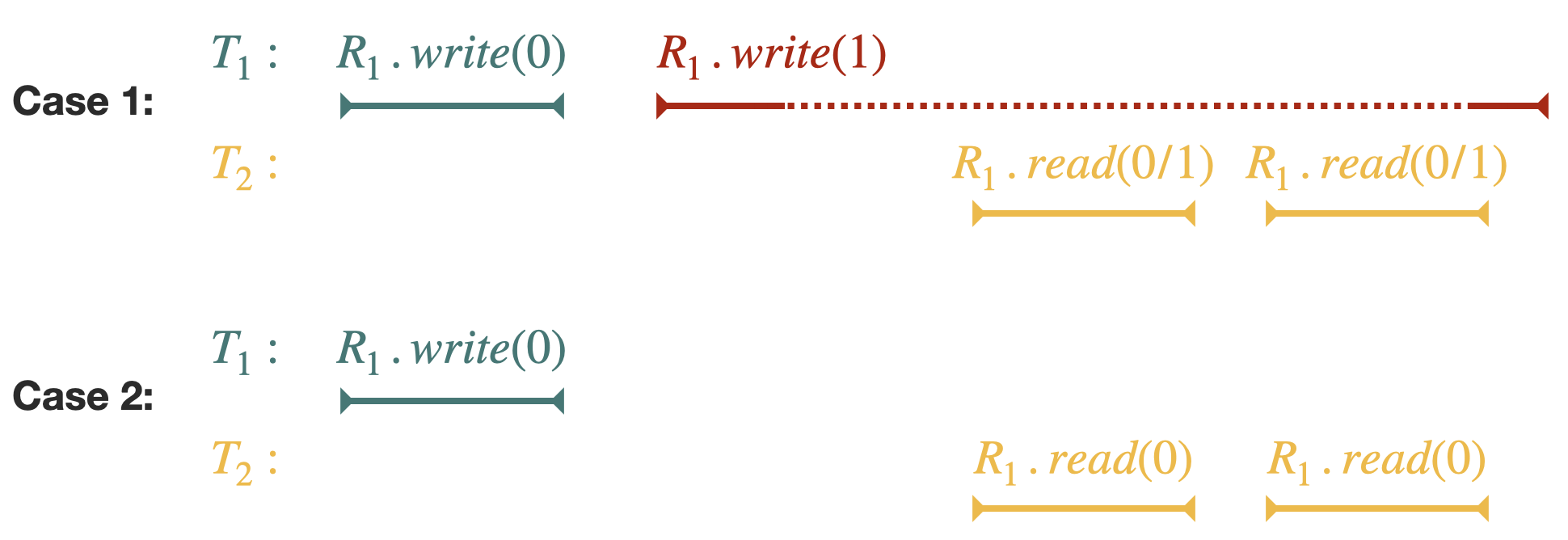}
    \caption{Durable history completions}
    \label{fig:durablecompletions}
\end{figure}

    An example is depicted in Figure~\ref{fig:durablecompletions}. This example matches the results shown for the linearizability definition, as expected. In contrast to linearizability, the notion of the durable linearizability definition adds crash events to the history and requires the history to be linearizable when crash events are removed from the completed history. 

    \textbf{Buffered Durable Linearizability}, was also defined by Izraelevitz et al.~\cite{izraelevitz2016linearizability}. This definition is more relaxed than durable linearizability in the sense that it may remove operations that have already completed, as long as it makes a `reasonable' history. Following~\cite{izraelevitz2016linearizability}, a history $H$ can be partitioned by crashes in the following manner: $H = H_0c_1H_1c_2..H_{c-1}c_tH_t$, where there exist $t$ crash events in $H$.  
    Formally, a history $H$ is \emph{buffered durably linearizable} if $H\vert p$ is sequential for every thread $p$ when crashes are removed from their histories, and for every sub-history $H_i$ between two consecutive crashes, there exists a $<_H$-consistent cut of $H_i$, denoted as $H_{s_i}$, such that $H_s = H_{s_0}H_{s_1}...H_{s_t}$ is linearizable.
    
\begin{figure}[htp]
    \centering
    \includegraphics[width=10cm]{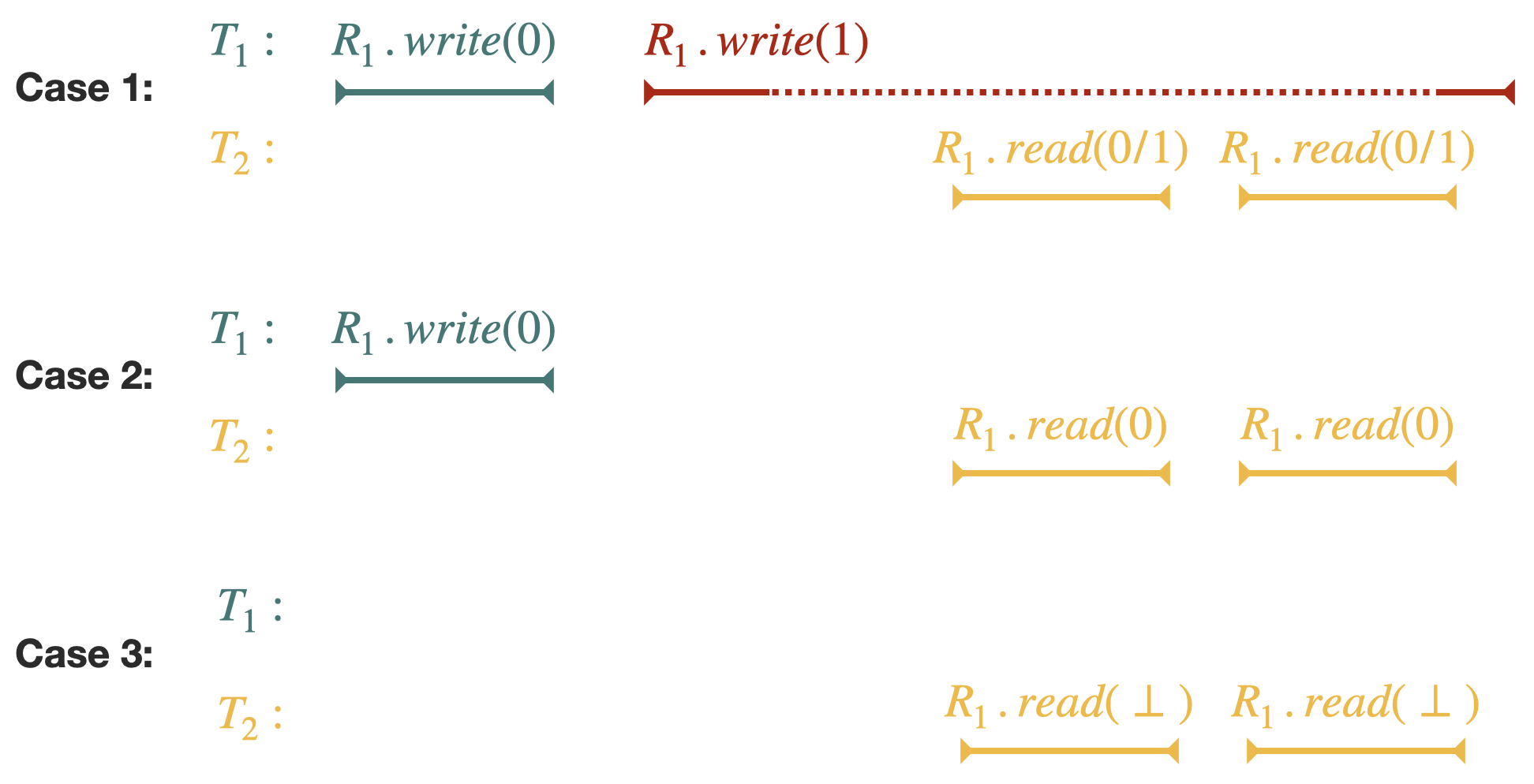}
    \caption{Buffered durable history completions}
    \label{fig:buffereddurablecompletions}
\end{figure}

    All the histories that satisfy durable linearizability also satisfy buffered durable linearizability. However, Buffered durable linearizability is not local~\cite{izraelevitz2016linearizability}. Additionally, removing all executed operations preceding the crash also satisfies buffered durable linearizability. Figure~\ref{fig:buffereddurablecompletions} demonstrates the possible histories according to our example. 
    Case $3$ represents a legal history where all executed operation before the crash were removed. Therefore, the reads by $T_2$ of $R_1$ return the initial value of the register, before anything has been written to it.

\subsection{New Definitions}

We note that the definitions above vary by when they allowed an interrupted operation is allowed to complete. They all focused on the placement of the response of an operation relative to the next invocation by the same process. Another way to define persistence correctness could be allowing the placement of a response relative to the next invocation by any process \emph{on the same object}. Therefore, we define two new properties, called \emph{social} and \emph{conservative} linearizability, based on this notion. We then show that they are both equivalent to strict linearizability.
 
    \textbf{Conservative Linearizability}
    A \emph{conservative completion}, $H'$, of a history $H$ can be obtained from $H$ by, for each pending invocation $inv$, either removing $inv$ from the history, or adding a matching response before the next invocation that accesses the same object or the next invocation by the same process as $inv$ after the crash event that interrupted $inv$ (whichever comes first).
    Conservative completions are indeed quite conservative, in that other than strict completions, they require the earliest placement of matching responses in a history.
    
    We thus define a history $H$ to be \emph{conservatively linearizable} if there exists a conservative completion of $H$ that satisfies conditions~\ref{cond:lin1} and \ref{cond:lin2} from the linearizability definition. 
    
    \textbf{Social Linearizability}
    
    We adapt a similar definition to Berryhill et al.~\cite{berryhill2016robust} for invoked before, but only relating to the same object (instead of same process and same object).
     Given a history $H$ and two operations $op_1$ and $op_2$ invoked on the same object $O$, $op_1$ is \emph{relaxed invoked before} $op_2$ (denoted by $op_1\lll_Hop_2$) if $inv$ of $op_1$ precedes $inv$ of $op_2$ in $H$.

    A history $H$ satisfies social linearizability if there exists a \emph{conservative completion} $H'$ and there exists a legal sequential history $S$ such that:
     \begin{enumerate}
         \item \label{cond:solin1} $H'$ is equivalent to a legal sequential history $S$
         \item \label{cond:solin2} If $op_1$ happens before $op_2$ in $H$, then $op_1$ happens before $op_2$ in $S$
         \item \label{cond:solin3} If $op_1$ is relaxed invoked before $op_2$ in $H$, then $op_1$ happens before $op_2$ in $S$
     \end{enumerate}
    
    The name \emph{social} linearizability refers to the fact that, unlike in persistent and recoverable linearizability, a process might be responsible to help an operation that is not its own complete, if it intends to use the same object.

    These definitions may relate to individual crashes and same processes are allowed to re-execute after crash. We relate to these possibilities later.

\begin{figure}[htp]
    \centering
    \includegraphics[width=10cm]{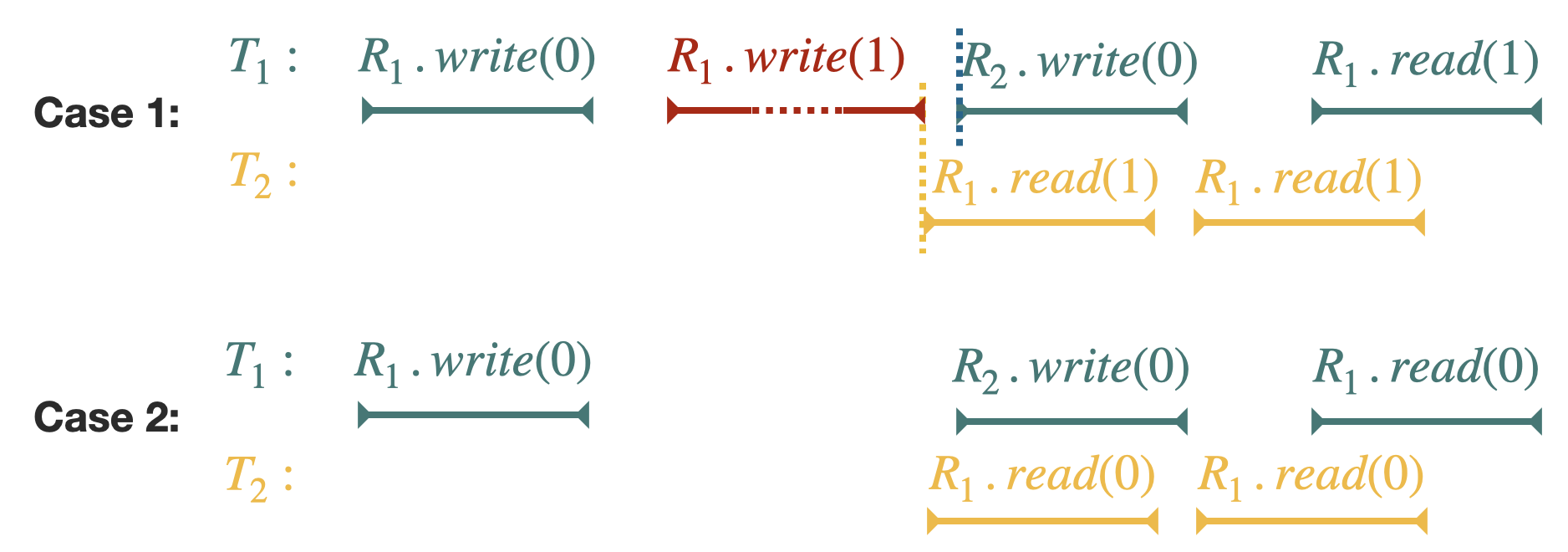}
    \caption{Social and Conservative histories}
    \label{fig:sociableconservativecompletions}
\end{figure}   

    The example representing histories that satisfy both executions is depicted in Figure~\ref{fig:sociableconservativecompletions}. Social linearizability requires ending the operation before the next invocation of the same object, as shown in case $1$. Therefore, from the moment that $T_2$ reads from $R_1$, it always returns the value $1$. As $T_2$ accesses $R_1$ before any invocation executed by $T_1$ in our example, case $1$ represents a legal history for the conservative linearizability as well. The second case simple remove the crashed operation from the history. Therefore, reads from $R_1$ must always return the value $0$ in our example.

    Accesses to the same object may seem a bit tricky. Requiring to add a matching response to pending invocations before the next invocation that accesses the same object $O$ is equivalent to strict linearizability since we can construct an equivalent history which either removes a pending invocation or adds a matching response right before the crash. Therefore, both conservative and social linearizability are in fact equivalent to strict linearizability.
    
    If, however, other objects might be influenced by the recovery of $O$, without actually accessing $O$, then these definitions become different from each other and from strict linearizability. For example, using a log which is not considered as part of the object and can indicate whether the object was recovered can be useful to differentiate between the definitions. However, such a log cannot exist in the standard concurrency model -- this would circumvent the definition of an object, and is generally not allowed.

\subsection{Continuing Executions}
The definitions so far focused on how to preserve the effects of operations on memory despite system crashes. They focused on capturing which operations' effects must be preserved and which ones may be lost for a persistent implementation to be considered correct. Furthermore, all of the definitions above have disallowed partial effects of operations to be persisted, therefore maintaining memory consistency.

However, there is another important issue when considering crash recovery. In particular, even if the memory is in a consistent state and preserved the effects of the right operations, this may be useless to a program if it cannot continue its execution where it left off. For example, imagine a producer-consumer program, in which one process has a list of tasks that it enqueues onto a shared queue, and another process dequeues and executes those tasks. After a system crash, if the shared queue was designed to satisfy one of the correctness definitions in the previous subsection, the queue will be in a consistent state, containing exactly the tasks that had been dequeued but not enqueued by some point in the execution before the crash. However, now the enqueuing process must continue its execution from where it left off. How does it know what part of its list of tasks it had already enqueued? The queue might not hold that information -- for example, it may be empty. In this case, the enqueuing process has no way of continuing its execution without running the risk of repeating or skipping tasks.

To address the above issue, several definitions have been proposed. These definitions focus on how to know where in the program a process should recover. They can also be combined with one of the definitions in the previous subsection.

    \textbf{Detectability.} Friedman et al.~\cite{friedman2018persistent} were the first to make the observation that in many programs, it is important to know whether a given operation interrupted by a crash has taken effect when recovering from that crash. An object is said to be \emph{detectable} if it provides a mechanism that can tell, for every operation that was executed while the crash occurred, whether or not that operation was completed. Detectability can therefore be added to any of the definitions in the previous section as an extra requirement for implementations to satisfy.

    \textbf{Nesting-Safe Recoverable Linearizability (NRL)} was defined by Attiya et al. to address the issue of continuing higher-level executions in the calling program after a crash occurs~\cite{attiya2018nesting}. Since programs are not usually designed for the possibility that an operation they call might not be executed (due to a potential crash), NRL requires that the most nested operation be completed immediately upon recovery from a crash, by executing a special \emph{recovery event} that is operation-specific. The paper originally defining NRL~\cite{attiya2018nesting} is concerned with how to implement concurrent objects that allow for such recovery.
    The NRL model specifies more assumptions to allow such implementations.
    In particular, the model assumes that all shared objects are non-volatile, processes can crash individually, and that the program counter, as well as the calling stack frame that contains a response value of an operation, are not lost upon a crash. In practice, these assumptions are difficult to satisfy.
    
    

    \textbf{Capsules.} Similarly to NRL, \emph{capsules}~\cite{blelloch2018parallel,ben2019delay} represent another way to ensure an implementation can recover operations that were interrupted by a crash. Capsules are a way of breaking a program into contiguous idempotent parts~\cite{blelloch2018parallel}. The idea is that each part, or \emph{capsule}, can be safely repeated from the beginning any number of times. Thus, at the boundary between two capsules, all variables are persisted, allowing a recovered process to continue its execution from the beginning of the most recent capsule it reached in the program. Capsules thus provide a method of implementing a persistent program. Just like in the NRL model, the capsules model also assumes that processes can crash individually and all shared objects are non-volatile. However, it is more flexible than NRL in its requirements on the implementation, since the stack frame and program counter need not be persisted on every instruction. 
    
    \textbf{Detectable Sequential Specifications (DSS)} shift the definition of detectability from history completions to sequential specifications~\cite{li2021detectable}. A detectable version $D<S>$ of a sequential specification $S$ adds two new operations for every operation $op$ in $S$; $prep-op$ and $exec-op$. Intuitively, to make a specific instance of $op$ detectable, $prep-op$ must be called first to notify the system that it should remember the outcome of the upcoming operation, followed by $exec-op$ to actually execute the operation. After $prep-op$ is called, we say that $op$ has been \emph{prepared}. Another new operation, $resolve$ can them be called after a crash, and returns the outcome of the most recently prepared operation. DSS also allows the option to simply call $op$ without preparing it, in cases in which that particular instance does not need to be detectable. By specifying detectability on sequential specifications in this way, DSS can be paired with any of the definitions in the previous section to make a detectable version of those properties. However, it does put an extra burden on the developer, who must implement three times as many operations as the non-detectable sequential specification. 
    \Michal{What if there is no outcome of the most recently prepared operation? Does it just indicates that the operation has not been completed?} \Naama{Is DSS the same as capsules and NRL when interpreted as history completions? Should we add it to the heirarchy?}

  \subsection{Mutual Exclusion}  
    \textbf{RME.} Another relevant definition was presented by Golab et al.~\cite{golab2019recoverable} to deal with mutual exclusion under executions with crashes. RME~\cite{golab2019recoverable} assumes that after every crash event, threads call a recovery procedure which cleans up the lock, tries to re-enter the critical section and when successful, re-executes the critical section from the beginning.
    A crucial difference between RME and the earlier definitions in this section is that RME does not ensure correctness of the overall program in the presence of crashes, it just ensures that standard mutual exclusion properties~\cite{lamport2019mutual} (i.e. mutual exclusion, deadlock-freedom/starvation-freedom, terminating exit/wait-free exit) hold in any execution without an infinite number of crash/recovery events. 
    To support nested critical sections, they also define the Critical Section Re-entry (CSR) and Bounded Critical Section Re-entry (BCSR) properties which ensure that if a process $p$ crashes within a critical section, then no other process can enter that critical section until process $p$ recovers and re-enters it.

\section{Property Hierarchy}

In this section, we present  hierarchies relating the properties presented in the previous section to each other under various model assumptions. Using different model assumptions regarding if and when crashed processes must recover influences the definitions hierarchy. Note that there is no difference in the hierarchy when individual crashes are replaced with system-wide crashes.

Recall that the two new definitions, social and conservative linearizability, are equivalent to strict linearizability. This is true under all versions of the model, so in our discussion comparing the different properties, we focus on strict.

\subsection{Same Processes are Invoked}

\begin{figure}[htp]
    \centering
    \includegraphics[width=10cm]{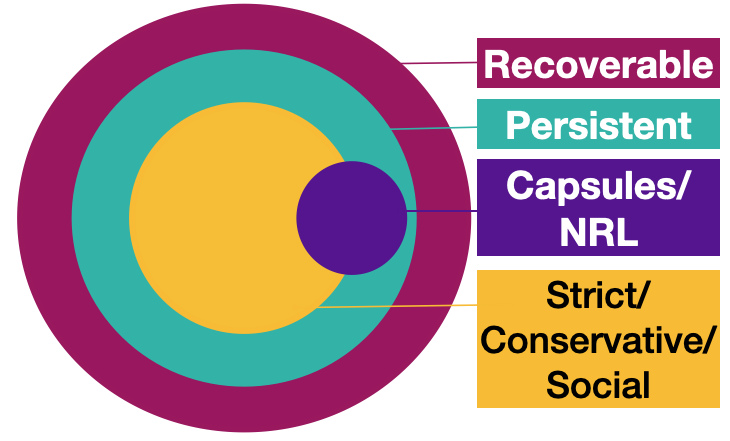}
    \caption{Hierarchy of definitions when the same threads are allowed to be invoked after a crash.}
    \label{fig:samethreads}
\end{figure}


In this subsection, we assume that the model allows the same processes to be invoked after a crash. Under this model, the definitions can be arranged into the hierarchy that is presented in Figure~\ref{fig:samethreads}.
The hierarchy is based on the sets of execution histories that are allowed by each of the definitions; in Figure~\ref{fig:samethreads}, each definition's set of allowed histories is represented by its labelled region. The diagram is not meant to represent the size of the sets, but only their relationship to each other. 

Note that durable linearizability is not defined in this model. 
It therefore does not appear in the hierarchy. We note that the buffered property can be applied not just to durable linearizability, but to all other correctness conditions as well~\cite{izraelevitz2016linearizability}. However, we omit it from the hierarchy in this subsection in the interest of clarity.

To understand this hierarchy, it is useful to consider how each correctness condition allows linearizing a given history. The correctness conditions differ by where they allow each pending operation's completion to be placed.
Berryhill et al.~\cite{berryhill2016robust} presented recoverable, persistent, and strict linearizability in this light. 

Recall that in a history $H$ in a non-volatile memory model, an operation $op$ by a process $p$ is considered \emph{pending} if its invocation, $inv_{op}$ is not followed by a matching response. This may occur for two reasons; either $inv_{op}$ was the last step of $p$ in the history, or $inv_{op}$ is the last step of $p$ before a crash event (or both).
There are several points in a given history with respect to which it may make sense to complete such a pending operation. One point of reference is the crash event that immediately follows $inv_{op}$ in $H|p$. Another is the next invocation by $p$ in the history. Finally, we may also consider the next invocation in $H$ that occurs in the same object as $op$.

To present the relationships in this hierarchy, we begin with a discussion that excludes capsule correctness and NRL, since their specification is incomparable to many others. At the end, we show how these correctness conditions can fit in with the rest.


Strict linearizability~\cite{aguilera2003strict} is the strongest (or \emph{strictest}) condition, in that it allows for the smallest set of histories. It requires every pending operation to be eliminated or completed before the crash. In addition, it is local, meaning that every object that is built from strictly linearizable objects is also strictly linearizable. To achieve this guarantee, one may think of running a recovery operation directly after the crash, and before re-executing the program. However, it might be too restrictive; in some scenarios, it makes sense to relax this requirement to allow recovery (alternatively; the completion of pending operations) to occur later in the execution. Furthermore, strict linearizability is known to disallow some surprising constructions; wait-free implementations of multi-reader single-writer registers from single-reader single-writer registers are not possible in a strictly linearizable manner~\cite{aguilera2003strict, berryhill2016robust}.

\begin{figure}[htp]
    \centering
    \includegraphics[width=10cm]{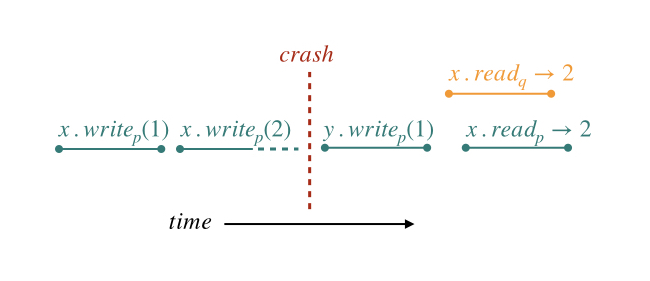}
    \caption{Example for showing non-locality of two objects and two executing threads. A crash occurs in the middle.}
    \label{fig:examplelocality}
\end{figure}

While strict linearizability requires completions to be placed before the next crash event, persistent atomicity~\cite{guerraoui2004robust} instead completes operations before the next invocation by the same process. Note that, by the definition of legal histories, the next invocation by the same process can never be placed before the current crash, and therefore persistent atomicity is weaker than strict linearizability (i.e. the set of persistent histories \emph{contains} the set of strict histories). 
Due to this relaxation, it is not local. 
An example by~\cite{berryhill2016robust} is presented in Figure~\ref{fig:examplelocality}. 
On the positive side, persistent atomicity may be easier to implement than strict linearizability, since an operation only needs to be recovered (if ever) when the same process invokes another operation.
\Michal{Should we update it to be more coherent with our updated example?}

Berryhill et al.~\cite{berryhill2016robust} presented the recoverable linearizability definition, which is more relaxed than the other two. It also requires pending operations to be completed (or removed) before the crash, but practically, to `take effect' before the next invocation of the same thread on the same object.
We now show that persistent atomicity is strictly stronger than recoverable linearizability.
To see why, suppose $H'$ is a persistent completion of $H$ that satisfies conditions 1 and 2 from the linearizability definition.
$H'$ is equivalent to a legal sequential history $S$ by condition 1 and we know that $S$ preserves the happens before relation of $H'$ by condition 2.
$H'$ preserves the happens before relations in $H$ by construction, so by transitivity, $S$ preserves the happens before relations of $H$.
The same is true for the invoked before relationship because $S$ is equivalent to $H'$ and $H'$ does not rearrange any of the invocations in $H$.
Therefore to show that $H$ satisfies recoverable linearizability, it suffices to construct a strict completion $H_{s}$ such that $H_{s}$ and $S$ are equivalent.
This can be done by starting with $H'$, and for each response in $H'$ that was not in $H$, we move it to before the preceding crash event from $H$. This would yield a strict completion which is equivalent to $S$, as required, since the relative order of operations on each process has not changed. 

We've now covered all definitions that were stated as requirements on history completions. However, to get the full picture, we now consider two more definitions, that were defined differently and for a different purpose. In particular, we consider NRL~\cite{attiya2018nesting} and Capsules~\cite{ben2019delay}. Recall that both of these definitions are concerned with allowing executions to continue after a crash, and do not allow any operation to be removed from a history. 

That is, unlike previous correctness conditions, a pending invocation may not be removed from a history.  
Thus, we can interpret them in the light of history completions, to better compare them to the other definitions, as follows.
An \emph{NRL (or Capsule) completion}, $H'$, of a history $H$ must, for every pending invocation $inv$, contain a matching response before the next invocation by the same process. For a history $H$ to be \emph{nesting-safe recoverably linearizable} (or capsule linearizable), it must have an NRL (or Capsule) completion that satisfies conditions~\ref{cond:lin1} and \ref{cond:lin2} from the linearizability definition. Under this interpretation, NRL and Capsules are equivalent.
However, we note that, as explained above, this view considers only which histories are allowable, but loses details on how information is stored and how an implementation can complete its operations, which differentiates NRL and Capsules under their original models.

\begin{figure}[htp]
    \centering
    \includegraphics[width=10cm]{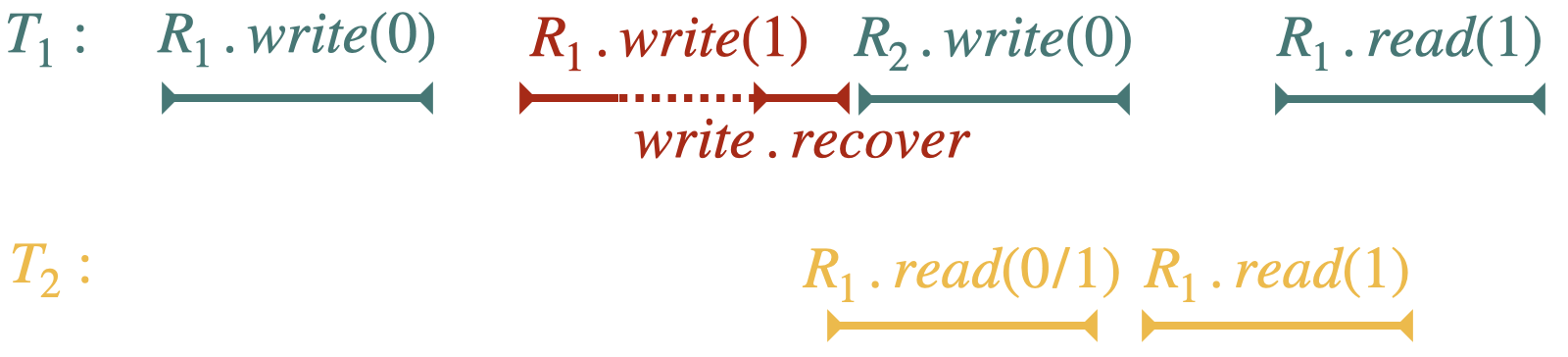}
    \caption{Nesting safe history completions}
    \label{fig:nestingsafecompletions}
\end{figure}

We present our running example with NRL/Capsules in Figure~\ref{fig:nestingsafecompletions}. The interrupted write operation must recover before the next invocation by the same process. Therefore, all subsequent reads from $R_1$ that are invoked after the recovery terminates, must return the value $1$.

Thus, with this history-based interpretation of these definitions, no definition we discussed so far implies NRL/Capsules. However, NRL and Capsles do overlap with the other definitions. In particular,  any strictly linearizable or persistent atomic completion in which no operation was removed is also consistent with NRL/Capsules. However, since NRL/Capsules only require completion of an operation before the next one by the same thread, we can construct an NRL/Capsule consistent history in which some  thread $t$ executes an entire operation after the crash before a thread $t'$ completes its operation which was started before the crash. Such a history is not strictly linearizable. However, it is still persistently atomic. In fact, persistent atomicity is strictly stronger than NRL/Capsules under this history completions interpretation.

Following these definitions, we can observe that there are two types of histories that seem to be missing, according to the timing of the completion of pending operations. As noted above, this is relevant only if there are other objects that might be influenced by the crashed object before actually accessing it.

First is the conservative linearizable definition that was first defined in this survey. It looks as if it relaxes the strict linearizability definition as it allows the pending operations to be also completed after the crash but before the next invocation of the pending process or the next invocation on the same object as the object of the pending operation. It also satisfies locality. Here, the recovery might be lazier, meaning that only if the same objects/processes are ever used, they will first recover the crashed operation. 

The second missing definition is social linearizability that has a similar approach to persistent atomicity~\cite{guerraoui2004robust}, but from the objects' perspective. Instead of completing the pending operations before the next invocation of the same process, it completes the pending operation over an object, before the next invocation of an operation on the same object. By changing the approach from processes to objects, we again get locality. Here, every object may have a recovery operation which may ran lazily only when there is another operation after the crash which is executed over that object. Note that histories which contain consecutive invocations may also satisfy this definition.


\subsection{New processes are invoked}

\begin{figure}[htp]
    \centering
    \includegraphics[width=10cm]{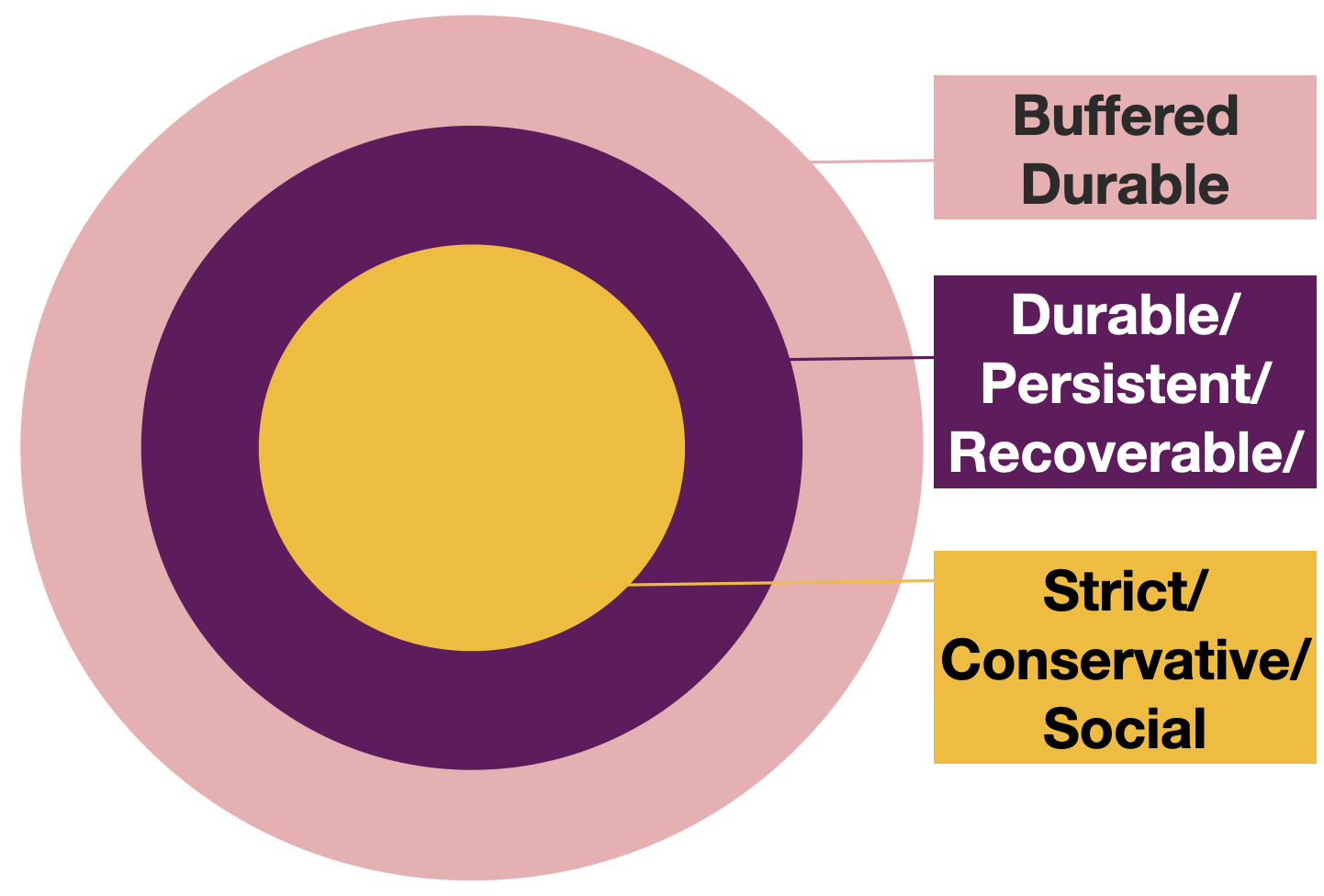}
    \caption{Hierarchy of definitions when only new threads are allowed to be invoked after a crash}
    \label{fig:individualcrashes}
\end{figure}

In this section, we assume that the model does not allow the same processes to be invoked after a crash, and new processes are spawned instead. This model was first suggested by Izraelevitz et al.~\cite{izraelevitz2016linearizability}, as a simplification to previous models. Under this simplification, the definitions that deal with execution continuations do not make sense. We therefore omit capsules and NRL from the discussion of this model. 
The hierarchy in this model is presented in Figure~\ref{fig:individualcrashes}.



When the same processes are never invoked after a crash, strict linearizability~\cite{aguilera2003strict} still remains the strongest condition as it requires every pending operation to be eliminated or completed before the crash. Recall that the difference between persistent atomicity~\cite{guerraoui2004robust} and recoverable linearizability~\cite{berryhill2016robust} is only in recoveries by the same process, and requires completing operations before the next invocation by the same process, whereas recoverable linearizability~\cite{berryhill2016robust} relaxes this requirement by practically recovering before the next invocation by the same process accessing the crashed object. Since same threads are never allowed to be executed after a crash, a response of a crashed operation can be added at the end of the history (or remove the invocation). Thus,
these two definitions have the same meaning as durable linearizability~\cite{izraelevitz2016linearizability}, which requires getting a linearizable history after removing all crash events from the original history.

By disallowing the executions of the same processes, durable linearizability, persistent atomicity and recoverable linearizability are local under this restriction.
Buffered durable linearizability is similar to the others, but additionally allows operations that were completed before the crash to be removed. It therefore is the weakest definition. As mentioned before, buffered durable linearizability~\cite{izraelevitz2016linearizability} allows more histories than durable linearizability but not local. Even if two objects satisfy buffered durable linearizability, a history composed out of these sub-histories may not satisfy this condition, as there is not guarantee to which prefix of every sub-history actually recovers. To mitigate this problem, a \emph{sync} operation can be provided. This operation ensures that all previously ordered operations before the sync, have reached the persistent memory. This operation, however, is not part of the definition.

\bibliographystyle{plainurl}
\bibliography{biblio}

\end{document}